\documentstyle[11pt,proceed]{article}

\begin{document}
% next 8 lines are for the preprint only
\centerline{\null}
\vskip-0.8truein
\noindent
To appear in {\it Proceedings of the Workshop on High Throughput X-ray
Spectroscopy},
ed.\ by P. Sullivan \& H. Tananbaum
(Cambridge: Smithsonian Astrophysical Observatory), in press.
\vskip0.5truein

\title{High Resolution X-ray Spectra of Cluster Cooling Flows}

\author{Craig L. Sarazin}
\affil{Department of Astronomy, University of Virginia,
	Charlottesville, VA 22903-0818 U.S.A.}

\thispagestyle{empty}
\begin{abstract}
The cooling cores of clusters of galaxies are among the brightest thermal
X-ray line sources in the universe.
High resolution X-ray spectra would allow individual line fluxes to
be measured.
The fluxes of low ionization X-ray lines in cooling flows should be
directly proportional to the cooling rate at relatively cool temperatures,
and could be used to determine whether large amounts of gas are really
cooling out of the X-ray band (below $\sim$$10^6$ K).
By comparing the fluxes of many X-ray lines, we could
determine the amounts of gas at different temperatures and densities.
This would allow one to test the idea that cooling flows are highly
inhomogeneous, as required to explain their surface brightness profiles.

High resolution X-ray spectra can be used to determine velocities
in the hot gas in cooling flows.
If the gas is homogeneous, significant inflow velocities are expected.
However, if the gas is as inhomogeneous as we currently believe, the
inflow velocities are likely to be rather low ($v_r \la 30$ km s$^{-1}$).
The hot gas in cooling flows is probably strongly turbulent,
and turbulence is likely to be more important than either inflow or
thermal broadening in determining the shapes of lines.
Most of the central galaxies in cooling flows host radio sources,
and it appears that the radio lobes are expanding against, displacing,
and compressing the cooling X-ray gas.
Significant motions associated with the expansion of radio sources should
be detectable in X-ray lines.
If the cooling X-ray gas rotates and viscosity is not important, rotational
motions may also be detectable in X-ray line spectra.

Some resonance emission lines from cooling flows are expected to
be moderately optically thick, although the largest opacities occur for
(possibly unrealistic) homogeneous models without turbulence.
Resonant scattering can affect the profiles of these lines.
These lines will appear in absorption against a background X-ray continuum
source; 
either a background quasar or the active nucleus of the central cooling
flow galaxy can provide such a light source.
From the comparison of the same line seen in both absorption and emission,
one can derive an estimate of the distance to the cluster, independent of
the Hobble constant.
Finally, high resolution X-ray spectra of absorption edges associated with
the excess soft X-ray absorption seen toward some cooling flows can be used
to determine the redshift of the absorber, and decide whether it is
associated with the cooling flow or with our Galaxy.
\end{abstract}

\section{Introduction} \label{sec:intro}

Observations with the $Einstein$ Observatory and some earlier missions
established that large quantities of gas are cooling below
X--ray emitting temperatures in the cores of many clusters
(see Fabian et al.\ [1984, 1991], and
Fabian [1994] for reviews).
Typical cooling rates are
${\dot M}_{cool} \approx 100 \, M_\odot$ yr$^{-1}$.
The gas cools over a region whose typical radius is
$r_{cool} \sim 200$ kpc.
The cooling is due to the X-rays we observe.
The emission of X-rays is thermally significant but not very
dynamically significant;
the observations imply that the time scale for cooling $t_{cool}$
satisfies
\begin{equation} \label{eq:times}
t_{free-fall} \ll t_{cool} < t_{age} \, ,
\end{equation}
where $t_{free-fall}$ is the dynamical time scale, and $t_{age}$ is the
age of the cluster (or, at least, the time since the last major subcluster
merger).
This means that the inflow associated with the cooling is rather slow
(\S~\ref{sec:inflow}).

The X-ray evidence for the cooling gas comes from both
the X-ray images and X-ray spectra of cluster cores.
X-ray images of the central regions of clusters with cooling flows show
very strongly peaked surface brightnesses.
When one determines the electron densities and the cooling times implied
by these brightnesses, they reach values of $n_e \sim 0.1$ cm$^{-3}$
and $t_{cool} < 10^9$ yr.
In essentially every known case, there is a very bright elliptical
galaxy at the center of the cooling region.

X-ray spectra also indicate the presence of cooling gas.
Lower resolution spectra with ROSAT and ASCA show a radially
increasing temperature gradient in the central regions of cooling
flow clusters.
In non-cooling flow clusters, the central regions are nearly isothermal.
In many cases, the X-ray spectra of the central regions of cooling flow
clusters are not well-fit by any single temperature component, indicating
that a range of gas temperatures are present.

The most direct evidence of cooling comes from higher resolution X-ray
spectra which show low ionization X-ray lines.
As the gas in the center of the cluster cools and recombines, low
ionization X-ray lines are produced.
For example, the $Einstein$ Focal Plane Crystal Spectrometer (FPCS)
detected O~{\sc viii} and Fe~{\sc xvii} lines from the cooling cores
of the Virgo and Perseus cluster
(Canizares et al.\ 1988).
As I will discuss below (\S~\ref{sec:low_ion}), such lines provide
the most direct X-ray evidence that gas is cooling out of the X-ray
temperature band.

In this paper, I will discuss a few potential applications of high
resolution X-ray spectra to cluster cooling flows to illustrate what
might be done with the High Throughput X-ray Spectroscopy mission
(HTXS).
I will leave the discussion of the global properties of X-ray
cluster spectra and abundances in the able hands of Joel Bregman.
Also, I won't review the standard plasma diagnostics from X-ray
spectra, since they will also be reviewed by others.

\section{Individual X-ray Line Fluxes} \label{sec:lines}

\subsection{Low Ionization X-ray Lines} \label{sec:low_ion}

The greatest mystery about cooling flows is the nature of the
ultimate repository of the gas seen to cool through the X-ray band.
If cooling flows are long--lived phenomena,
the total amount of cooled gas is roughly
\begin{equation} \label{eq:mcool}
M_{cool} \approx 10^{12} \, M_\odot
\left( \frac{{\dot M}_{cool}}{100 \, M_\odot \, {\rm yr}^{-1}} \right)
\left( \frac{t_{age}}{10^{10} \, {\rm yr}} \right) \, .
\end{equation}
Although cooler material is seen in the optical and radio as
emission line gas, young stars, or cold gas,
the amount seen in other spectral bands is
considerably less than the total mass of cooled gas which is expected.
Thus, it is most important to confirm that the gas seen cooling through
the X-ray band does indeed cool to lower temperatures.

Theoretically, if thermal conduction is suppressed, radiative cooling
accelerates.
Once gas has cooled to a few million K, it is very difficult to
prevent it from cooling further.
However, it may be possible that some other mechanism intervenes to
reheat the gas back up to the ambient cluster temperature of
$\sim$$10^8$ K.
For example, Norman \& Meiksin (1996) have recently suggested that
reconnecting magnetic fields and thermal conduction might reheat the
cooling X-ray gas.

Measurements of the fluxes of individual low ionization X-ray lines
can allow one to determine directly the cooling rates of gas at low
temperatures.
For lines such as Fe~{\sc xvii} produced mainly at temperatures which
are much lower than the ambient cluster temperature, the gas can
be assumed to be cooling in steady-state, and the line luminosity
is given by
\begin{equation} \label{eq:linelum}
L_{line} = \frac52 \, \dot{M}_{cool} \, \frac{k}{\mu m_p}
\, \int_o^{T_o} \frac{\Lambda_{line} ( T )}{\Lambda_{tot} ( T )}
\, d T \, ,
\end{equation}
where
$\mu \approx 0.61$ is the mean mass per particle in terms of the proton
mass $m_p$,
$T_o$ is the temperature from which the gas starts to cool,
$T$ is its current temperature,
the emissivity of the line is $\rho^2 \Lambda_{line} ( T )$,
$\rho$ is the density of the gas, and the total emissivity
(cooling rate) of the gas is $\rho^2 \Lambda_{line} ( T )$
(e.g., White \& Sarazin 1987).
Typically, the emissivity of a low ionization line peaks at a
temperature $T_{max}$.
As long as $T_{max} \ll T_o$, the line luminosity is nearly independent of
most of the details of how the gas cools, and depends mainly on the
cooling rate through that temperature, $\dot{M}_{cool} ( T_{max} )$.
For example, the initial temperature of the gas, $T_o$ is not important.
The specific density at which the gas cools is also not important;
if the density
of the gas is increased, both the rate of line emission and the cooling
rate are increased by the same factor, and the amount of line emission per
gram of cooling gas remains unchanged.
Similarly, the line
luminosity is nearly independent of the overall abundance of heavy
elements.
This result occurs because the total cooling rate is dominated by
line emission by heavy elements at low temperatures, so changing the heavy
element abundance affects
$\Lambda_{line}$ and $\Lambda_{tot}$ by almost the same factor.
The specific numerical factor in front of equation~(\ref{eq:linelum})
assumes that the gas cools isobarically, which is essentially what is
expected, but the luminosity would only be reduced by $<$40\% is the gas
cooled more nearly isochorically.

Line fluxes from a number of individual low ionization X-ray lines can
therefore be used to determine the amount of cooling through a variety
of temperatures given by the values of $T_{max}$ for the lines.
If the gas is indeed cooling to low temperatures below the X-ray
band, then the implied cooling rates should all agree
[$\dot{M}_{cool} ( T_{max} ) = $ constant].
If the gas only cools part way through the X-ray band and then
is reheated, the implied values of $\dot{M}_{cool}$ for the lowest
ionization lines should be much lower than those derived from higher
ionization lines, or from other analyses of the X-ray spectrum or
surface brightness.

At present, the only data with sufficient spectral resolution to apply
this technique is still the $Einstein$ FPCS data, and the observations
of the Virgo and Perseus clusters are consistent with gas cooling out
of the X-ray band
(Canizares et al.\ 1988).
However, more observations of high resolution line spectra are
needed.

\subsection{Inhomogeneous Cooling Flow Gas} \label{sec:inhomo}

Beyond this determination of the amount of gas cooling out of
the X-ray band, the fluxes of X-ray lines can be used to
constrain the local density and temperature structure in the
cooling flow gas.
The simplest situation would be if the gas were locally homogeneous,
so that there was a single value of the gas density $\rho (r)$ and
temperature $T (r)$ at each radius $r$.
In homogeneous models, the gas doesn't cool to low temperature until
it flow into the center ($r \la 1$ kpc) of the cluster
(e.g., White \& Sarazin 1987).
On the other hand, the standard picture for cluster cooling flows is
that the gas is highly inhomogeneous, with a wide range of temperatures
and densities at each radius.
The different phases of gas are likely to be nearly isobaric, because
equation~(\ref{eq:times}) guarantees that the sound crossing time for
a clump of gas will be much less than its cooling time.
Thus, there are likely to be dense lumps of cool gas immersed in more
diffuse, hotter gas.
Now, gas which is either denser or cooler radiates more effectively
than gas which is hotter and less dense, so cooling amplifies
the temperature differences in the gas.
As a result, the cooler, denser lumps of gas will cool out of the
X-ray band at large radii, while the more diffuse, hotter gas will
not cool until it has reached the center of the flow.
In an inhomogeneous cooling flow, gas cools out of the X-ray band
over a very extended region of the order of 200 kpc in radius
(Thomas et al.\ 1987; White \& Sarazin 1987).

Analyses of the X-ray images of cluster cooling flows from the $Einstein$
Observatory and subsequently have indicated that the gas must be
quite inhomogeneous
(Thomas et al.\ 1987;
White \& Sarazin 1987).
The X-ray surface brightness profiles of cluster are strongly peaked
to the center, but not as strongly as expected if the gas were
homogeneous.
Most of the gas must cool below X-ray emitting temperatures at
relatively large radii $\sim$100 kpc.
Deconvolutions of the X-ray surface brightness profiles suggest that
the amount of gas cooling out of the X-ray band varies with radius
roughly as $\dot{M}_{cool} ( \le r ) \propto r$
(Thomas et al.\ 1987;
Fabian et al.\ 1991).

Nulsen (1986) and Thomas (1988) have developed the theory for
the density and temperature fluctuations in cluster cooling flows,
and some resulting X-ray spectra have been calculated by
Wise \& Sarazin (1993).
Let $f(T)$ be the temperature distribution function of gas,
defined such that $f(T) dT$ is the fraction of the total volume $V$
occupied by gas with a temperature between $T$ and $T + dT$.
Let us assume for the moment that the gas is isobaric, and let
$P$ be the pressure of the gas.
Then, the luminosity of any X-ray line is given by
\begin{equation} \label{eq:lum_fluct}
L_{line} =
\left( \frac{\mu m_p P}{k} \right)^2 \, V \,
\int f(T) \Lambda_{line} ( T )
\frac{d T}{T^2} \, .
\end{equation}
For steady-state cooling, at low temperatures $ T \ll T_o$ the
temperature distribution function approaches
\begin{equation} \label{eq:f_steady}
f_{steady} (T) = \frac52 \, \dot{M}_{cool} \,
\left( \frac{k}{\mu m_p} \right)^3 \,
\left( \frac{1}{P^3 V} \right) \,
\left[ \frac{T^2}{\Lambda_{tot} ( T )} \right] \, .
\end{equation}
Substituting equation~(\ref{eq:f_steady}) into equation~(\ref{eq:lum_fluct})
recovers equation~(\ref{eq:linelum}).

As noted in \S~\ref{sec:low_ion}, different X-ray lines vary with
temperature in different ways, and the emissivities of many lines 
(e.g., the Fe L lines) peak at some temperature $T_{max}$.
Thus, by comparing the fluxes in a variety of lines, one can determine
or constrain the distribution of density and temperature in the gas.
The spectrally derived distribution function can be compared to that
required to explain the X-ray surface brightness profiles.
Also, the distribution functions of densities and temperatures at the
outer edge of a cooling flow can be used to derive the degree of
density fluctuations in the general intracluster gas.
Among other things, this is needed to correct distances derived by
the comparison of the Sunyaev-Zel'dovich effect and the X-ray ray
emission of clusters for the effects of density fluctuations.

\section{Velocity Fields} \label{sec:velocities}

\subsection{Radial Inflow} \label{sec:inflow}

Of course, one of the main applications for high resolution
X-ray spectroscopy will be to use the Doppler effect to
measure velocities in the hot gas.
In the case of cooling flows,
it would be very exciting if one could detect the radial inflow
associated with the cooling and accretion of the X-ray emitting
gas by the central galaxy.
Unfortunately, the inflow in cooling flows is generally
very slow and subsonic (\S~\ref{sec:intro}),
and this will not be easy to do.

One can derive an approximate expression for the inflow velocity
$v_r$ in a cooling flow by noting that mass conservation requires that the
cooling rate $\dot{M}_{cool} (r)$ interior to $r$ must equal
the inflow rate $4 \pi r^2 \rho v_r $ in steady-state.
The density can be estimated by assuming that the velocity is
also given approximately by $v \approx r / t_{cool}$.
This leads to the following approximate expression for the
inflow velocity, which is found to agree reasonably well
with detailed hydrodynamical models:
\begin{equation} \label{eq:vradial}
v_r \approx 9 \,
\left( \frac{r}{100 \, {\rm kpc}} \right)^{-1/2} \,
\left[ \frac{\dot{M}_{cool} (r)}{300 \, M_\odot \, {\rm yr}^{-1}}
\right]^{1/2} \,
\left[ \frac{T(r)}{10^8 \, {\rm K}} \right]^{-1/2} \,
{\rm km} \, {\rm s}^{-1} \, .
\end{equation}

\begin{figure}[t]
\vspace{2.55in}
\includegraphics{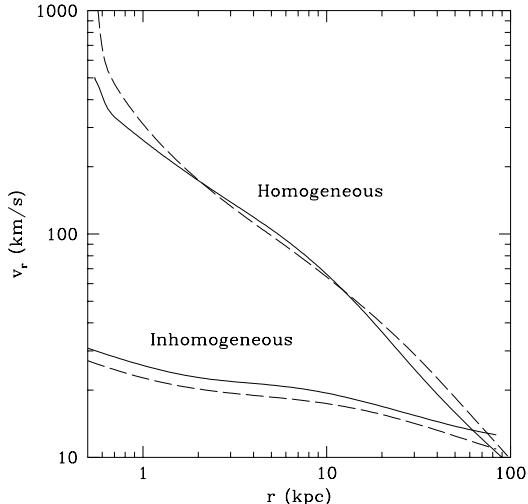}
\caption{The inflow velocity $v_r$ in cooling flow models.
In both of the models shown here, the total cooling rate is
300 $M_\odot$ yr$^{-1}$ and the ambient cluster gas has
a temperature of $8 \times 10^7$ K.
The models are taken from Wise \& Sarazin (1993).
The upper and lower sets of curves are for homogeneous
and highly inhomogeneous cooling flow models;
the inhomogeneous model has $\dot{M}_{cool} (r) \propto r$.
The solid curves are the actual values from the hydrodynamical
models, while the dashed curves are the velocities from
equation~(\protect\ref{eq:vradial}).}
\label{fig:inflow}
\end{figure}

The variation of the inflow velocity with radius is illustrated in
Figure~\ref{fig:inflow}.
The solid curves show the inflow velocities for detailed hydrodynamical models
from Wise \& Sarazin (1993), while the dashed curves are the result of using
the approximate expression in equation~(\ref{eq:vradial}).
For both of the models shown in Figure~\ref{fig:inflow}, the total cooling
rate is 300 $M_\odot$ yr$^{-1}$, and the ambient cluster gas temperature is
$T_o = 8 \times 10^7$ K.
The inflow velocities are always low ($\sim$10 km s$^{-1}$) at the outer
edge of the flow.
The upper curves in Figure~\ref{fig:inflow} are for a homogeneous model
in which the gas has a single density and temperature at each radius.
In such a homogeneous model, the gas remains hot until it reaches the
innermost part of the flow ($r \la 1$ kpc), where it cools very rapidly.
Because all of the gas which enters the cooling flow continues to flow into
the center, $\dot{M}_{cool} (r)$ is constant, and the inflow velocity grows
to a large value $v_r \sim 300$ km s$^{-1}$ at the center.

\begin{figure}[t]
\vspace{2.5in}
\includegraphics{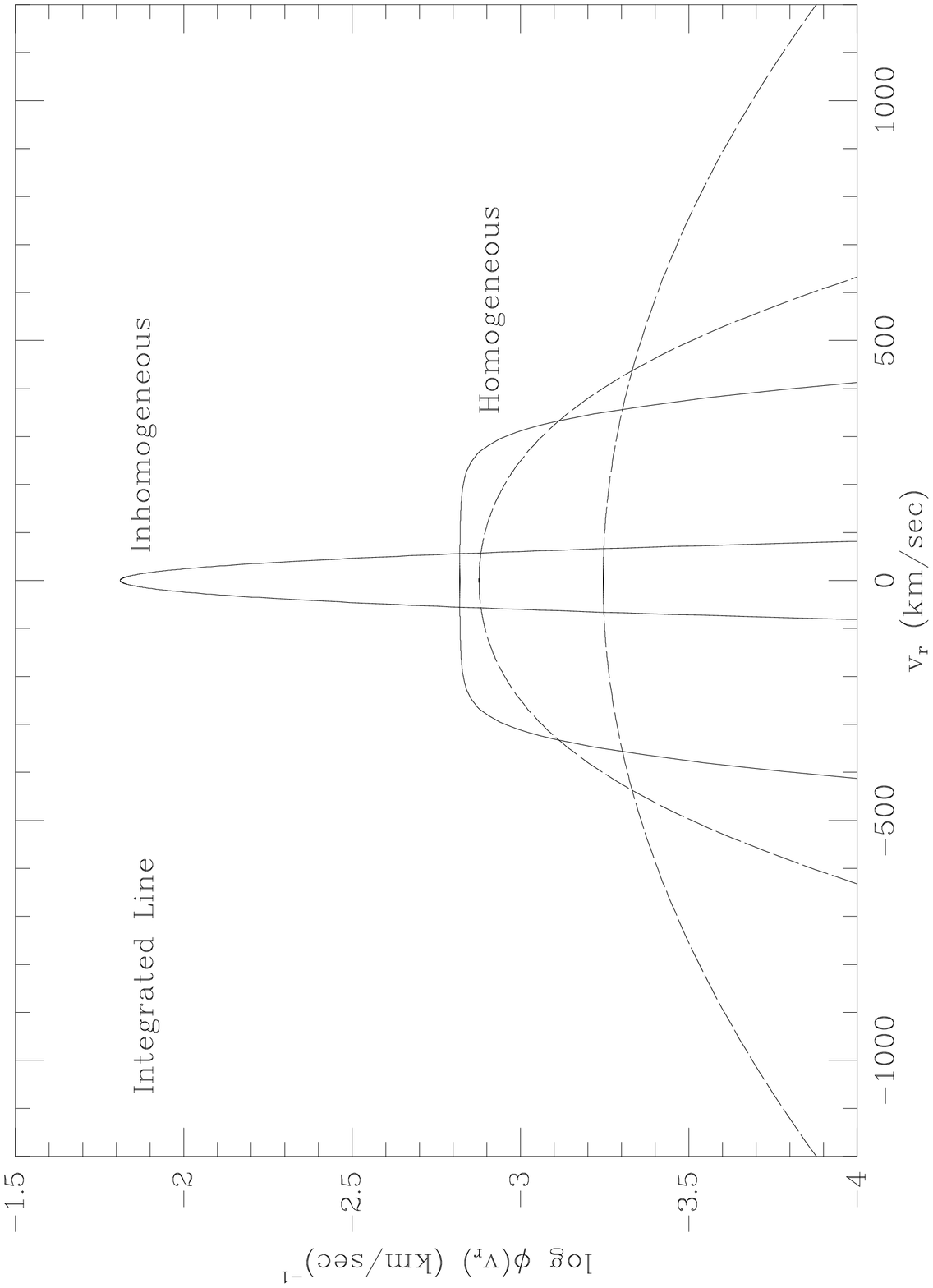}
\caption{The integrated line profiles for the Fe~{\sc xvii} line from
the entire cooling flow region in homogeneous and inhomogeneous
cooling flow models
(Wise \& Sarazin 1993).
The models are the same as those in Figure~\protect\ref{fig:inflow}.
The solid curves give the profiles assuming only thermal broadening.
The dashed curves are for maximal, transonic turbulent broadening.
}
\label{fig:profile_tot}
\end{figure}

In the inhomogeneous model, gas cools below X-ray emitting temperatures
throughout the center of the cluster.
In the specific inhomogeneous model shown Figure~\ref{fig:inflow},
$\dot{M}_{cool} (r) \propto r$, which agrees with the analyses of the
surface brightness profiles of many clusters
(e.g., Fabian et al.\ 1991).
Since the mass flux of inflowing gas declines towards the center,
the inflow velocity does not increase as rapidly as in the homogeneous case.
In fact, for $\dot{M}_{cool} (r) \propto r$, the effect of the radius term
and $\dot{M}$ term in equation~(\ref{eq:vradial}) cancel one another,
and the only reason the velocity increases in the center is
that the temperature drops there, which speeds up the cooling.
As discussed in \S~\ref{sec:inhomo}, the analysis of existing X-ray data
strongly favor inhomogeneous models.
In such models, the inflow velocities are generally small
($v_r \la 30$ km s$^{-1}$), which will make it quite difficult to detect
the inflow with X-ray spectroscopy.

\begin{figure}[t]
\vspace{2.5in}
\includegraphics{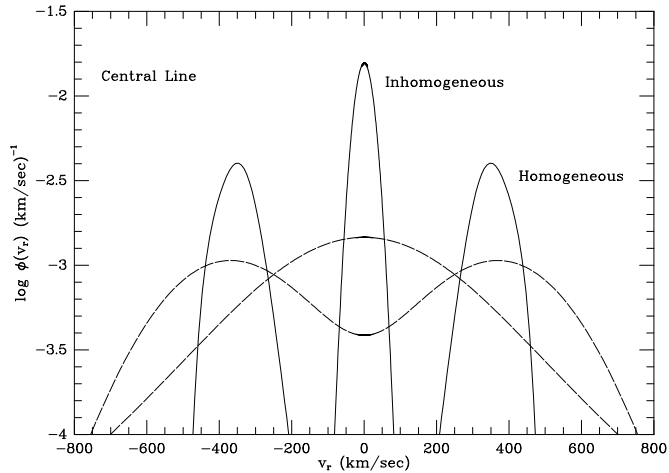}
\caption{The line profiles of the Fe~{\sc xvii} line for a line
of sight through the center of the cooling flow.
The models and notation are the same as in
Figure~\protect\ref{fig:profile_tot}.
Note that the thermally- broadened line from the homogeneous model
is double.}
\label{fig:profile_cnt}
\end{figure}

The effect of inflow velocities on line profiles is illustrated in
Figures~\ref{fig:profile_tot} and \ref{fig:profile_cnt}.
Figure~\ref{fig:profile_tot} shows integrated line profiles for
the Fe~{\sc xvii} line from the entire cooling flow region of a cluster
for homogeneous and inhomogeneous models.
The models are the same as those in Figure~\ref{fig:inflow}.
The solid curves assume only thermal line broadening, while the dashed
curves show the opposite extreme of transonic turbulent broadening
(\S~\ref{sec:turbulent}).
Figure~\ref{fig:profile_cnt} shows the same line profiles, but for
the center of the cooling flow (in projection).
In practice, these spectra would require a spatial resolution of
roughly 1 kpc, which translates into $\la$1 arcsec for typical cluster
cooling flow.
Thus, HTXS will have too little spatial resolution to see spectra
comparable to Figure~\ref{fig:profile_cnt} in all clusters except
possibly Virgo, and Figure~\ref{fig:profile_tot} is more relevant
to HTXS.

For a homogeneous cooling flow, the high velocities near the center
(coupled with the strong low ionization line emission from this region)
produce strong distortions in the shapes of low ionization X-ray emission
lines.
These lines have a very non-gaussian square shape in the integrated
spectrum (Figure~\ref{fig:profile_tot}), and are double in the
central spectrum (Figure~\ref{fig:profile_cnt}).
Turbulent broadening reduces these effects, but doesn't eliminate
them.
However, in the more realistic inhomogeneous models, the inflow velocities
are small and the line emission is distributed over the entire cooling flow
region.
Line profiles for strongly inhomogeneous models do not show any strong
effects of inflow, and are, in practice, indistinguishable from gaussians.

In conclusion, it will be difficult to measure inflow velocities
in cluster cooling flows with X-ray spectra if they are as inhomogeneous
(and turbulent) as is generally thought.

\subsection{Turbulent Motions} \label{sec:turbulent}

Most astrophysical gases are turbulent, and one seldom encounters line
profiles from heavy elements which are dominated by thermal broadening.
It is likely that the intracluster gas in cooling flows is strongly
turbulent.
We already have indirect evidence for this from the profiles of the
optical emission line nebulae which are observed almost universally
in the central $\sim$20 kpc of cluster cooling flows
(e.g., Baum 1992).
The optical emission lines are quite broad, and imply turbulent velocities
of 200 -- 1000 km s$^{-1}$.
It is likely that the emission line gas is embedded in the X-ray emitting
gas, and that the X-ray gas has similar velocities.
The upper limit on the turbulent velocity of the hot gas is presumably
the sound speed, since supersonic turbulent motions would damp quickly.
This is consistent with the largest values of the turbulent velocities
from the emission lines.
Thus, it is likely that X-ray emission lines from cooling flows
will be broadened by turbulent velocities of
\begin{equation} \label{eq:vturb}
v_{turb} \approx 200 - 1000 \, {\rm km} \, {\rm s}^{-1} \, .
\end{equation}

In Figures~\ref{fig:profile_tot} and ~\ref{fig:profile_cnt}, the
broad line profiles (dashed curves) show the effects of maximal,
transonic turbulence on the profiles of X-ray lines.
It seems likely that turbulence will dominate over thermal broadening
and radial inflow in determining the profiles of X-ray lines from
cooling flows, at least if cooling flows are strongly inhomogeneous.
Of course, the term ``turbulence'' often refers to complex unresolved
motions of any origin, rather than real hydrodynamical turbulence.
There is some evidence for interesting structure due to radial
velocity stratification or occasionally rotation in the kinematic of
the optical emission line nebulae of some cooling flows
(e.g., Baum 1992).
These motions may arise from interesting physical processes, and may lead
to more interesting line profiles or spatial variations than homogeneous,
isotropic turbulence.
I consider two illustrative examples: the expansion of radio lobes,
and disks due to rotating gas.

\subsection{Radio Lobe Expansion} \label{sec:radio}

In nearly every large cluster cooling flow, the central galaxy hosts
a moderately strong radio source.
Conversely, many of the most famous radio sources in the nearby universe
lie at the centers of cooling flows
(Virgo A, Perseus A, Cygnus A, Hydra A, \dots).
These are generally fairly compact, FR~I radio sources.
X-ray observations with ROSAT have provided evidence for the interaction
of the lobes of these radio sources with the ambient X-ray emitting
gas in cluster cooling flows
(Sarazin 1996).
First, the nonthermal pressures of the radio lobes inferred from
synchrotron theory are generally comparable to or slightly greater
than the thermal pressures of the X-ray emitting gas at the same
projected radii.
Second, the radio images and X-ray images are anticorrelated,
in the sense that the radio lobes are faint in X-rays.
The best example of this is the Perseus cluster
(B\"ohringer et al.\ 1993), but a number of other cases have been
found
(Huang \& Sarazin 1996; Sarazin 1996).
In Perseus, it appears that the X-ray emission is concentrated just
outside the edges of the radio lobes.
Third, the radio lobes in cooling flows are generally polarized
but have extremely large Faraday rotations
(Taylor et al.\ 1994).
This implies that there is very little thermal plasma within the
radio lobes themselves.
All of this is consistent with the idea that the radio plasma has expanded
against and displaced the X-ray emitting gas.

In some cases, one also finds that the optical emission line gas is
concentrated around the edges of the radio lobes
(Baum 1992).
In a number of cases, McNamara \& O'Connell (1993) found regions
of blue optical continuum emission at the edges of radio lobes
in cooling flows.
In A1795, HST observations have resolved these blue lobes into
knots of young stars
(McNamara et al.\ 1996).
A simple explanation for these phenomena is that the radio lobes
are compressing the surrounding X-ray emitting gas, which causes
it to cool more rapidly and to form stars.

From the moderate increase in the surface brightness of the X-ray
emitting gas at the edges of the radio lobes and the fact that
the radio pressures are similar to but somewhat higher than the
X-ray pressures, it seems likely that the expansion of the
radio lobes (or other motions) are mildly supersonic with respect
to the X-ray gas.
Under these conditions, the velocities induced in the X-ray gas
might be of the order of
\begin{equation} \label{eq:vradio}
v_{radio} \approx 300 - 1000 \, {\rm km} \, {\rm s}^{-1} \, .
\end{equation}
Thus, these motions should be detectable in high resolution
X-ray spectra with HTXS.

\subsection{Accretion Disks?} \label{sec:disk}

In the previous section, I noted the anticorrelation which
exists between the radio and X-ray images of the inner regions
of cluster cooling flows.
In some cases, the X-ray emission appears to be elongated perpendicular
to the radio axis.
One explanation of this, discussed above, is that the radio lobes
displace the X-ray emitting gas.
Another possible explanation is that the X-ray gas in the inner regions
forms a flattened rotating configuration, and that this rotating gas
eventually feeds an accretion disk around the AGN
(Sarazin et al.\ 1995).

Although clusters do not rotate rapidly, the intracluster gas presumably
has some small amount of angular momentum.
As the gas cools and contracts into the center of the cooling flow,
its rotation speed will increase if the gas preserves it angular
momentum.
Alternatively, 
Nulsen et al.\ (1984)
have argued that turbulent viscosity will effectively transport
the angular momentum of cooling flow gas outward.
However, if the gas does spin up as it flows inward, it will
eventually form a rotating disk.
(Mathews discussed galactic equivalents to this in his talk
at this meeting.)
Such a disk would rotate with a typical speed of
\begin{equation} \label{eq:vrot}
v_{rot} \approx 300 \, {\rm km} \, {\rm s}^{-1} \, ,
\end{equation}
and would produce X-ray emission lines with the classical
double horned profile of a rotating disk.
These might be detectable in X-ray spectra with HTXS, although
they have not generally been seen in optical emission line
profiles from cooling flows.

\section{Opacity Effects} \label{sec:opacity}

\subsection{Optically Thick Emission Lines} \label{sec:optthick}

Cluster cooling flows contain large columns of moderately hot gas,
and under some circumstances, resonant X-ray emission lines can
be optically thick
(Gil'fanov et al.\ 1987).
Wise \& Sarazin (1996) have calculated the spectra of cooling flows
including the radiative transfer in the lines.
The optical depths of the lines can reach values as high as
$\tau \sim 10^2$, although the opacities are greatest in
homogeneous cooling flow models without turbulence.
As noted above (\S\S~\ref{sec:inhomo}, \ref{sec:inflow},
\ref{sec:turbulent}), these models are probably not very
realistic.

\begin{figure}[t]
\vspace{2.65in}
\includegraphics{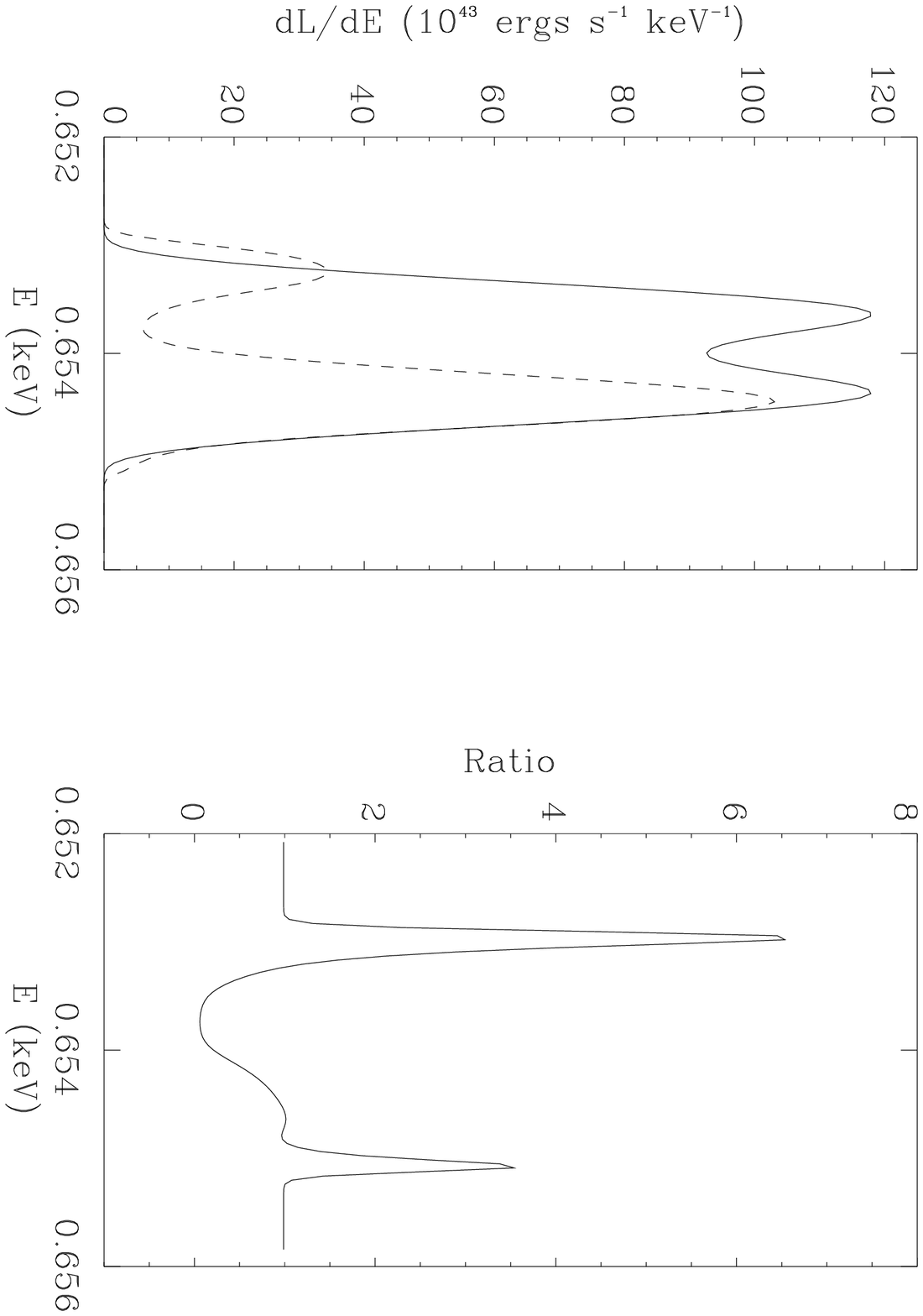}
\caption{The effects of resonant scattering on the O~{\sc viii}
line seen toward the center of a homogeneous cooling flow
model without turbulence
(Wise \& Sarazin 1993).
The solid curves give the profile with no line opacity,
while the dashed curves include the line optical depth.}
\label{fig:thickprofile}
\end{figure}

The opacity in these lines is due to resonant scattering.
In the absence of any source of absorptive opacity
(see \S~\ref{sec:excess}), this scattering doesn't remove
any photons from the lines.
Instead, it redistributes photons spatially and in wavelength.
This causes resonant emission lines to appear to come from further
out in the cluster,  and it changes the line profiles of the
scattered photons.
Figure~\ref{fig:thickprofile} illustrates this effect for the
O~{\sc viii} line toward the center of a homogeneous cooling flow
model without turbulence
(Wise \& Sarazin 1996).
There is a reduction in the red wing of the line, which arises in
gas flowing toward the center on the near side.

\subsection{Absorption Lines} \label{sec:absorb}

These same resonant lines would appear as absorption lines in the
spectra of a bright continuum source lying behind the cluster.
Obviously, a background quasar or other AGN might provide such
a light source.
As noted in \S~\ref{sec:radio}, the central galaxies in cluster
cooling flows are often radio galaxies, and in some cases these
AGN are bright in X-rays.
A particularly interesting case is A1030, where the central
cluster galaxy hosts the radio quasar B2 1028+313.
An X-ray bright AGN in the center of a cluster cooling flows
provides the ideal geometry for detecting absorption lines.

Krolik \& Raymond (1988) and Sarazin (1989)
showed that the detection of an absorption
line toward a background quasar, accompanied by the detection of
the same line in emission, would allow a direct determination of
the distance to the cluster.
This technique is very similar to the application of the
Sunyaev--Zel'dovich effect.
Crudely, the strength of the absorption is independent of
the distance to the quasar, while the flux of the emission
line declines with the square of the distance.
Comparing the two provides an estimate of the distance.
While initially discussed for absorption in the spectrum
of a quasar behind any region of a cluster, the same technique
could be applied to an absorption lines in the spectrum of
a central AGN due to the surrounding cooling flow.
The ratio of emission flux to absorption optical depth also
depends on the density of the gas.
It may be that this technique will be more useful for determining
the fluctuations in density of the gas, rather than as a tool
in cosmology.

\subsection{Excess X-ray Absorption} \label{sec:excess}

From a reanalysis of the $Einstein$ Solid State Spectrometer
(SSS) spectra of clusters of galaxies,
White et al.\ (1991) concluded that the spectra of cooling flows
require considerable amounts of soft X-ray absorption in excess
of the columns through our Galaxy.
The excess columns were found to be
$\Delta N_H \approx 10^{20} - 4 \times 10^{21}$ cm$^{-2}$.
The SSS spectra generally covered only the central cooling flow
region of the cluster ($r \sim 200$ kpc);
also, in several cases, the ROSAT PSPC has shown that the excess absorption
is concentrated to the cluster center, $r \sim 200$ kpc
(Allen et al.\ 1993;
Irwin \& Sarazin 1995).
If the extra absorber is associated with the cluster, then the excess
column and area imply a total mass of absorbing material of
$M_{cold} \sim 10^{11} - 3 \times 10^{12} \, M_\odot$.
This is quite similar to the total mass of gas ($M_{cool}$)
expected to cool out during the lifetime of a cluster
(equation~\ref{eq:mcool}).
This excess absorption is very important, in that it represents the only
direct evidence which has been found for a reservoir for the cooled gas
which could contain the total amount of this material.

Of course, high resolution X-ray spectra are not needed to detect excess soft
X-ray absorption.
Observations with the ROSAT PSPC have confirmed the presence of such
absorption in a number of cases
(Allen et al.\ 1993, 1995;
Irwin \& Sarazin 1995),
but there are also a number of strong cooling flows that do not show
much evidence for excess absorption
(e.g., Sarazin \& McNamara 1996).
It is obviously of great importance to determine whether this excess
absorption is really associated with the cluster cooling flow, as opposed
to being a feature in the interstellar medium of our own Galaxy.

At the columns of interest, the K-edge of oxygen is the dominant spectral
feature in the absorption, followed by the edges of nearby elements.
High resolution X-ray spectra could be used to measure the wavelength
and shape of the absorption edge.
If the emission is due to gas at the cluster, the edge will be redshifted
(Wise \& Sarazin 1996).
Once the redshift is established, measuring the precise wavelength and
shape of an absorption edge can provide information on the ionization state
of the absorber.
It is currently assumed that the absorption is due to cold atomic or
molecular gas or dust grains.

If the excess absorption is associated with the cluster cooling flow, it
provides a source of continuum absorption within the flow.
As such, it would be particularly effective at removing optically thick
resonant emission lines, which must scatter a number of times before
escaping the cluster
(\S~\ref{sec:optthick}; Wise \& Sarazin 1996).

\acknowledgments

This work was supported by NASA Astrophysical Theory Program grant
NAG 5-3057, NASA ASCA grant NAG 5-2526, and NASA ROSAT grant 5-3308.


\begin{references}
 
\reference{}
Allen, S. W., Fabian, A. C., Johnstone, R. M., White, D. A., Daines, S. J.,
Edge, A. C., \& Stewart, G. C. 1993, MNRAS, 262, 901
 
\reference{}
Allen, S. W., Fabian, A. C., Edge, A. C., B\"ohringer, H., \& White, D. A.
1995, MNRAS, 275, 741

\reference{}
Baum, S. A. 1992,
in Clusters and Superclusters of Galaxies, ed.\ A. C. Fabian
(Dordrecht: Kluwer), 171

\reference{}
B\"ohringer, H., Voges, W., Fabian, A. C., Edge, A. C., \& Neumann, D. M.
1993, MNRAS, 264, L25

\reference{}
Canizares, C. R., Markert, T. H., \& Donahue, M. E. 1988,
in Cooling Flows in Clusters and Galaxies,
ed.\ A. C. Fabian (Dordrecht: Kluwer), 63

\reference{}
Fabian, A. C. 1994, ARA\&A, 32, 277

\reference{}
Fabian, A. C., Nulsen, P. E., \& Canizares, C. R. 1984,
Nature, 310, 733

\reference{}
Fabian, A. C., Nulsen, P. E., \& Canizares, C. R. 1991,
A\&AR, 2, 191

\reference{}
Gil'fanov, M. R., Sunyaev, R. A., \& Churazov, E. M. 1987,
SovA, 13, L233

\reference{}
Huang, Z., \& Sarazin, C. L. 1996, preprint

\reference{}
Irwin, J. A., \& Sarazin, C. L. 1995, ApJ, 455, 497

\reference{}
Krolik, J. H., \& Raymond, J. C. 1988, ApJ, 335, L39

\reference{}
McNamara, B. R., \& O'Connell, R. W. 1993,
AJ, 105, 417

\reference{}
McNamara, B. R., Wise, M. W., Sarazin, C. L., Jannuzi, B. T., \& Elston, R.
1996, ApJ, 466, L9

\reference{}
Norman, C., \& Meiksin, A. 1996, ApJ, 468, 97

\reference{}
Nulsen, P. E. 1986, MNRAS, 221, 377

\reference{}
Nulsen, P. E., Stewart, G. C., \& Fabian, A. C. 1984, MNRAS, 208, 185

\reference{}
Sarazin, C. L. 1989, ApJ, 345, 12

\reference{}
Sarazin, C. L. 1996, in R\"ontgenstrahlung from the Universe,
ed.\ H. U. Zimmermann, J. E. Tr\"umper, \& H. Yorke (Garching: MPE), 561

\reference{}
Sarazin, C. L., Burns, J., Roettiger, K., \& McNamara, B. R.
1995, ApJ, 447, 559

\reference{}
Sarazin, C. L., \& McNamara, B. R.  1996, ApJ, submitted

\reference{}
Taylor, G. B., Barton, E. J., \& Ge, J. P. 1994, AJ, 107, 1942

\reference{}
Thomas, P. A. 1988, MNRAS, 235, 315

\reference{}
Thomas, P. A., Fabian, A. C., \& Nulsen, P. E. J. 1987,
MNRAS, 228, 973

\reference{}
White, D. A., Fabian, A. C., Johnstone, R. M., Mushotzky, R. F., \& Arnaud,
K. A. 1991, MNRAS, 252, 72

\reference{}
White, R. E., III, \& Sarazin, C. L. 1987, ApJ, 318, 612

\reference{}
Wise, M. W., \& Sarazin, C. L. 1993, ApJ, 415, 58

\reference{}
Wise, M. W., \& Sarazin, C. L. 1996, preprint

\end{references}
\end{document}